%% file: var7-gamma3-english.tex
\documentclass[a4paper,12pt]{article}
\usepackage[vmargin={25mm,25mm},hmargin={30mm,20mm}]{geometry}
\usepackage[T2A]{fontenc}
\usepackage[cp1251]{inputenc}
\usepackage[russian,english]{babel}
\usepackage{babelbib}
\usepackage[bookmarks=true,unicode=true]{hyperref}
\hypersetup{%
	pdftitle={Moon's perigee mass as a missing component of the Earth's precession-nutation theory},%
	pdfauthor={Kiryan~D.\,G., Kiryan~G.\,V.},%
	pdfkeywords={ASTROMETRY. LATITUDE. LONGITUDE. CHANDLER'S MOTION.
                            GRAVITATION. MOON. RADIO-ASTROMETRY. GPS. UT1.}%
}
\usepackage{siunitx}
\usepackage{amsmath,amsfonts,amssymb}
\usepackage{mathtext}
\usepackage[uline]{hhtensor}
\usepackage[stable]{footmisc}
\usepackage{eulervm}
\usepackage{graphicx}
\usepackage{marvosym}
\usepackage{misccorr}
\usepackage{commath} 
\usepackage{setspace}  
\usepackage{nicefrac}
\usepackage{url}
\usepackage{cancel}
\usepackage{xspace}
\usepackage[small]{caption2}
\usepackage[draft]{prelim2e}

\graphicspath{{images/}}

\def\g{\ensuremath{\mathbf{g}}\xspace}
\def\G{\ensuremath{\mathbf{G}}\xspace}

\begin{document}

\vspace*{10mm}
\centerline{\LARGE\textbf{Moon's perigee mass as a missing component of}}
\smallskip
\centerline{\LARGE\textbf{the Earth's precession-nutation theory}}
\bigskip
\medskip

\centerline{\large D.\,G.~Kiryan\footnote{Institute of Problems of Mechanical Engineering of RAS,
61 Bolshoy Prospect V.O., 199178,\\ Saint-Petersburg, Russia,
e-mail: \textit{diki.ipme@gmail.com}}, G.\,V.~Kiryan}
\vspace*{-10mm}

\input{var7-gamma3-english-abstract}

\input{var7-gamma3-english-history}

\input{var7-gamma3-english-problem}

\input{var7-gamma3-english-A}
\input{var7-gamma3-english-step}
\input{var7-gamma3-english-Luna}
\input{var7-gamma3-english-plumb-line}

\input{var7-gamma3-english-comparison}

\input{var7-gamma3-english-IERS}
\input{var7-gamma3-english-ABCD}
\input{var7-gamma3-english-conclusion}

\begin{spacing}{0.90}
    \bibliography{../../../_x-BIB/var,../../../_x-BIB/kiryan}
    \bibliographystyle{gost780u}
\end{spacing}

\end{document}

%% file: var7-gamma3-english-abstract.tex
%

\section*{}
\label{sec:abstract}

\begin{spacing}{0.85}
\noindent
\textit{In this work, the nutation momentum acting upon the Earth from the Moon's perigee mass that has not been taken into account in the Earth's precession-nutation theory was revealed. This missing momentum exhibits itself in the so-called "local latitude variation" with the Chandler's period. The results of our work raise the question of updating the Earth's precession-nutation theory and revising some postulates of the time service, astronomy, geophysics, satellite navigation, etc.}
\end{spacing}
\bigskip

\noindent
\textbf{Keywords:} astrometry, latitude, longitude, Chandler's wobble, gravitation, Moon, 
                radio-astrometry, GPS, UT1.


%% file: var7-gamma3-english-history.tex
%

\section{Observation of star zenith distances}

In astrometry, the zenith distance is defined as the angle between the plumb line (local normal) and direction to the star. It is commonly recognized that instability of star zenith distances was for the first time noted by J.\,Bradley (1726--1727) and Molyneux (1727--1747)~\cite{1972:book:Kulikov, 1982:book:Podobed}. In 1840, H.I.\,Peters was the first who purposefully detected the zenith distance variations (latitude variability) by using advanced optical instruments at The Central Astronomical Observatory of the Russian Academy of Sciences at Pulkovo. Similar observations were being performed at the same observatory from 1863 to 1875 by M.O.\,Nuren; he was the first who estimated the latitude variation period as $1.2$~year. The issue of giving to these investigations international character was discussed at the International Geodetic Association Congress in 1883 in~Rome. Practical observations were begun after the Salzburg Geodetic Association Congress (1888). In 1892, S.\,Chandler who has studied and generalized the observations acquired by that time showed that among the latitude variation periods there is one of $400$ to $440$~days~\cite{1902:AJ-v22-p145:Chandler}. At that moment, the fact that this phenomenon is caused by motion of the Earth's rotation axis within the Earth has already been regarded as evident. Soon enough, the following version of the phenomenon interpretation was offered to the scientific community: the star zenith distance variation (latitude variation) is caused by "free nutation motion" of the Earth's rotation axis within the Earth~\cite{1892:Newcomb}. This was the first attempt to explain the physical nature of regular variations in the star zenith distance.

Then, at the turn of the 20th century, the scientific community formulated the following hypothesis based on the Chandler's discovery and hypothesis on the ''latitude variation'' nature (that, in our opinion, has not been experimentally confirmed to the necessary and sufficient extent): the Earth's rotation axis performs within the Earth "residual motion" with the characteristic Chandler's period.

During the 20th century, other hypotheses were suggested; their common feature was that they were based on evident but not proved facts. What was assumed to be evident was that the observed variations in the star zenith distances were caused by the Earth's rotation axis displacement with respect the Earth's body. At the end of the 20th century, in the paper devoted to the Chandler's discovery 100th anniversary~\cite{1987:SGeo-v9-p419:Runcorn}, top-level European and American experts in the Earth's poles (rotation axis) motion and Earth's rotation theory had to state absolute absence of results of this phenomenon investigation.

This fact stimulated us to approach to solving this problem departing from the paradigm that has existed so long. Some findings of our studies in this field are presented in \cite{2003:BOOK:var1, 2012:CONFERENCE:APM:var7-gamma}. In the studies, specific attention was paid to the problem definition, necessary and sufficient conditions imposed on the observation procedure, requirements for the angular astrometric instruments used in investigation, and issues of retaining physical entity in analyzing the time series obtained.


%% file: var7-gamma3-english-problem.tex
%

\section{Problem definition}
\label{sec:problem}

Based on that the Chandler's wobble has been revealed in analyzing long-term observations of the ocean level variations~\cite{1952:article:Maximov}, atmospheric pressure~\cite{2009:article:Bogdanov}, and the Earth's gravitational acceleration\footnote{Ocean Hemisphere Project Data Management Center ---  \url{http://ohp-ju.eri.u-tokyo.ac.jp}}, we assumed that the Chandler's wobble nature is determined by a certain gravitational factor. Thus, let us consider the gravity field variation at a certain Earth's surface point as a function of the Moon's position with respect to the Earth by the example of the plumb line. The plumb line is the line tangential to the gravity field line at the observation point.

\begin{figure}[htb]
        \centering\includegraphics[scale=1]{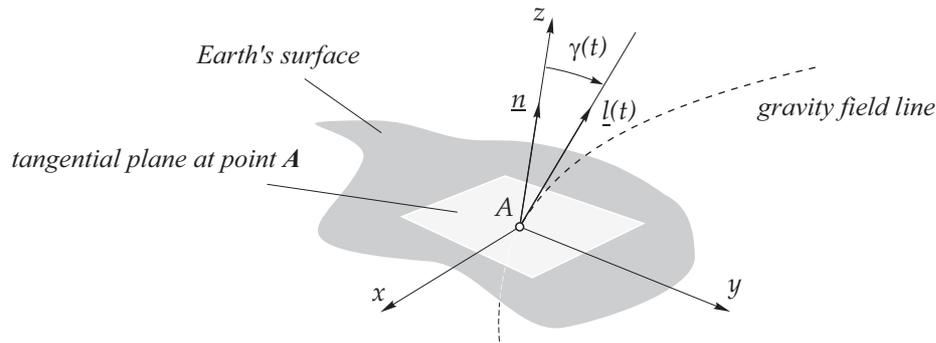}
        \caption{Angle of the gravity field line departure at the Earth's surface point~$\mathbf{A}$.}
        \label{fig:var7-gamma3-english-A-gamma}
     \end{figure}

Consider the gravity field instability at point $\mathbf{A}$ on the Earth's surface as a consequence of the Moon's motion about the Earth. Assume that the Earth and the Moon are rigid spheres. Let us include into consideration the orthogonal coordinate system $\mathbf{A}xyz$ with the origin at point $\mathbf{A}$ (Fig.~\ref{fig:var7-gamma3-english-A-gamma}). The coordinate system is oriented so that the $\mathbf{A}xy$ plane is tangential to the Earth's surface at point~$\mathbf{A}$. The~$\mathbf{A}z$ axis is directed towards the~Earth.

The force acting upon the proofmass at point $\mathbf{A}$ may be defined as the gradient of the
Earth-Moon system gravity field potential~$\mathbf{U}(t)$:
    \begin{equation}\label{eq:var7:gradU}
        \vec{f}(t)=-\nabla\mathbf{U}(t)\;.
    \end{equation}

By definition, force $\vec{f}(t)$ is tangential to the gravity field line. Let us follow variations in the force $\vec{f}(t)$ direction in the $\mathbf{A}xyz$ coordinate system via vector $\vec{l}(t)$ by calculating
angle $\gamma(t)$ between the fixed unit vector~$\vec{n}$ (axis $\mathbf{A}z$~ort) and vector~$\vec{l}(t)$ via the cosine theorem:
    \begin{equation}\label{eq:var7:gamma}
        \gamma(t)=acos\bigl(\: \vec{n}\cdot\vec{l}(t) \:\bigr)\;,\quad\text{where}\quad
        \vec{l}(t) = -\frac{\vec{f}(t)}{|\vec{f}(t)|}\;,\quad
        |\vec{l}(t)|=1\;.
    \end{equation}

We do not consider other physical phenomena causing the gravity field variations at the observation point; as it is shown below, they have no fundamental importance within the scope of our problem definition. This means that, being aimed at revealing true nature of the process, we have simplified the problem as far as possible; here we do not divert our attention to minor factors that, however, can play a significant role under other circumstances.


%% file: var7-gamma3-english-A.tex
%

\section{Observer's coordinates at the Earth's surface}
\label{sec:A}

\begin{figure}[htb]
        \centering\includegraphics[scale=1]{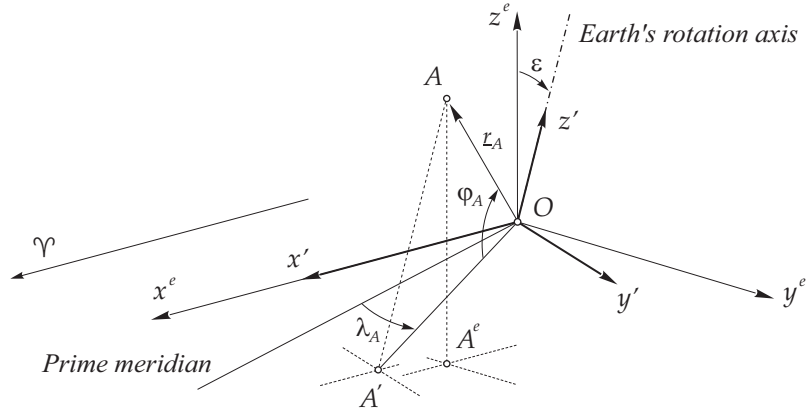}
        \caption{Observer's coordinates at the Earth's surface (point $\mathbf{A}$)}
        \label{fig:var7-gamma3-english-A}
     \end{figure}

Consider two fixed orthogonal coordinate systems $\mathbf{O}x^ey^ez^e$ and $\mathbf{O}x'y'z'$ with the common origin at point $\mathbf{O}$ (Fig.~\ref{fig:var7-gamma3-english-A}). Plane $\mathbf{O}x^ey^e$ belongs to the ecliptic, while plane~$\mathbf{O}x'y'$ coincides with the Earth's equator plane. Assume that the Earth is an ellipsoid of revolution, $\mathbf{O}z'$ is the axis of the Earth's self-rotation about maximal momentum of inertia, $\mathbf{O}z'$ makes angle~$\varepsilon$ with the $\mathbf{O}z^e$ axis, axes $\mathbf{O}x'$ and $\mathbf{O}x^e$ are of the same direction and are parallel~to~\Aries.

The point $\mathbf{A}$ coordinates on the Earth's surface are defined by the latitude and longitude.
Latitude $\varphi_A$ is the angle between the $\mathbf{O}x'y'$ plane (the Equator) and direction towards point~$\mathbf{A}$. Longitude $\lambda_A$ is defined as the angle in the $\mathbf{O}x'y'$ plane between the Prime meridian and point $\mathbf{A}$ meridian. The Prime meridian and point~$\mathbf{A}$ rotate as a unit about the $\mathbf{O}z'$ axis. Designate as $R_A$ the distance between the Earth's center of gravity $\mathbf{O}$ and point~$\mathbf{A}$. This distance depends on the latitude and parameters of the Earth (the ellipsoid of revolution):
    \begin{equation}\label{eq:var7:RA}
            R_A =R_A (\varphi_A, e_{terra}, a_{terra})\;.
    \end{equation}

\begin{figure}[htb]
        \centering\includegraphics[scale=1]{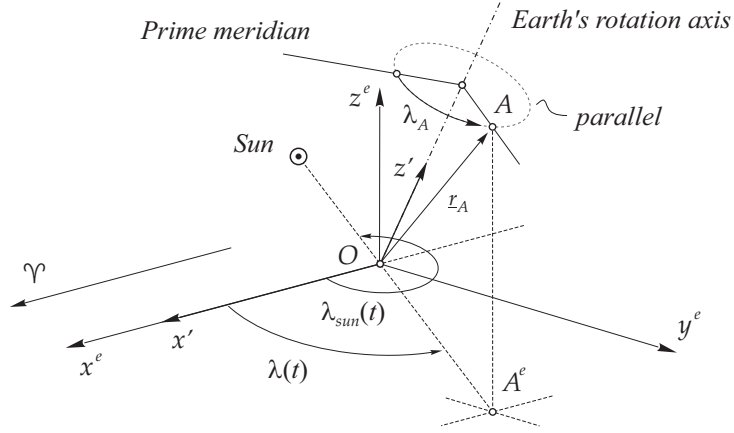}
        \caption{Realization of the "local midnight" event at observation point $\mathbf{A}$}
        \label{fig:var7-gamma3-english-Lambda}
     \end{figure}

In the fixed coordinate system $\mathbf{O}x^ey^ez^e$ (Fig.~\ref{fig:var7-gamma3-english-Lambda}), point $\mathbf{A}$ is defined by vector~$\vec{r}_A$:
    \begin{equation}\label{eq:var7:rA}
        \vec{r}_A(t)=R_A\!\cdot\!
            \mathbf{P}_x\bigl(\varepsilon\bigr)\!\cdot\!
            \mathbf{P}_z\bigl(\lambda(t)\bigr)\!\cdot\!
            \begin{pmatrix}
                    \cos\varphi_A \cos\lambda_A \\
                    \cos\varphi_A \sin\lambda_A \\
                    \sin\varphi_A \\
            \end{pmatrix}\;,\quad |\vec{r}_A(t)|=R_A\;,
    \end{equation}
where $\mathbf{P}_x$, $\mathbf{P}_z$ are the rotation matrices\footnote{\input{var7-gamma3-english-footnote-PxPyPz}}; $\mathbf{A}^e$ is the point $\mathbf{A}$ projection on the ecliptic plane $\mathbf{O}x^ey^e$; $\lambda(t)$ is the angle in the ecliptic plane $\mathbf{O}x^ey^e$ between axis $\mathbf{O}x^e$ and line passing through points $\mathbf{O}$ and~$\mathbf{A}^e$. Define angle $\lambda(t)$ via the Sun longitude $\lambda_{sun}(t)$ for the relevant epoch~\cite{1989:BOOK:AY1991}. For instance, at the time moment~$t$ when the so-called "local midnight" takes place at point~$\mathbf{A}$,
    \begin{equation}\label{eq:var7:Lambda}
            \lambda(t)=\lambda_{sun}(t) - \pi\;.
    \end{equation}

Note that term \emph{solar day}\footnote{The solar day is the time interval between two consecutive upper culminations of the Sun, which is 24~hours. Note that, being measured in angles, the solar~day exceeds~$2\pi$ because of the Earth's rotation about the Sun.} we are accustomed to stipulates two rotations: the Earth's self-rotation and annual rotation of the Earth about the Sun:
    \begin{equation}\label{eq:var7:OmegaTerra}
            \omega(t)=\omega_\ast(t) + \omega_\text{year}(t)\;,
    \end{equation}
where $\omega_\ast(t)$ is the Earth's self-rotation angular velocity (the Earth revolves about its axis with respect to stars during the time period of about $\!23^{h}56^{m}04^{s}$); $\omega_\text{year}(t)$ is the extra angular rotation velocity ensuring our customary sequence of sunrises and sunsets, namely, solar~day. Since the diurnal latitude measurements shell be reduced to one and the same time moment (e.g., "local midnight"), we reject the "fast" component of the Earth's daily rotation~$\omega_\ast$; hereinafter, the Earth's rotation velocity is presented only by the extra angular velocity~$\omega_\text{year}(t)$. This extra angular velocity is the time derivative of the Sun longitude~$\lambda_{sun}(t)$; hence, relation~\eqref{eq:var7:OmegaTerra} takes the following form:
    \begin{equation}\label{eq:var7:OmegaSun}
            \omega(t)=\cancelto{0}{\omega_\ast(t)}+\omega_\text{year}(t)=
            \od{\lambda_{sun}(t)}{t}\approx\frac{2\pi}{T_\text{year}}\;,\;\;\text{where}\quad
            T_\text{year}=\!365.25\text{\ days}.
    \end{equation}

This means that we assume the Earth to rotate about the Sun facing the Sun always by one and the same side. Hence, the Observer located at any selected point on the Earth's surface will retain his orientation relative to the direction towards the Sun. Hereinafter we assume the Earth's rotation angular velocity to be constant (time-independent).


%% file: var7-gamma3-english-footnote-PxPyPz.tex
%
Let point $M$ in the $\mathbf{O}xyz$ coordinate system be defined by vector $\vec{r}$; then, in the
new coordinate system $\mathbf{O}x'y'z'$ formed by turning $\mathbf{O}xyz$ by angle $\xi$ about axis $\mathbf{O}x$, point $M$ will be defined by vector $\vec{r}'$. Matrices of rotation by angle $\xi$ about axes $\mathbf{O}x$, $\mathbf{O}y$ and $\mathbf{O}z$ are given below. Counterclockwise rotation is assumed to be "positive".
\begin{equation}
    \vec{r}'=\mathbf{P}_x(\xi)\vec{r}\;,\quad
    \mathbf{P}_x(\xi)=\begin{pmatrix}
               \phantom{-}1 & \phantom{-}0               & \phantom{-}0\phantom{-} \\
               \phantom{-}0 & \phantom{-}\cos(\xi) &                   -\sin(\xi)\phantom{-} \\
               \phantom{-}0 & \phantom{-}\sin(\xi) & \phantom{-}\cos(\xi)\phantom{-}\\
             \end{pmatrix}\;,\quad
    \vec{r}=\mathbf{P}_x^{-1}(\xi)\vec{r}'\nonumber
 \end{equation}
\begin{equation}
    \vec{r}'=\mathbf{P}_y(\xi)\vec{r}\;,\quad
    \mathbf{P}_y(\xi)=\begin{pmatrix}
               \phantom{-}\cos(\xi) & \phantom{-}0 &                  -\sin(\xi)\phantom{-} \\
               \phantom{-}0              & \phantom{-}1 &                     0\\
               \phantom{-}\sin(\xi) & \phantom{-}0 & \phantom{-}\cos(\xi)\phantom{-}\\
             \end{pmatrix}\;,\quad
    \vec{r}=\mathbf{P}_y^{-1}(\xi)\vec{r}'\nonumber
 \end{equation}
\begin{equation}
    \vec{r}'=\mathbf{P}_z(\xi)\vec{r}\;,\quad
    \mathbf{P}_z(\xi)=\begin{pmatrix}
          \phantom{-}\cos(\xi) & \phantom{-}\sin(\xi)         & \phantom{-}0\phantom{-}\\
                             -\sin(\xi)  & \phantom{-}\cos(\xi)        & \phantom{-}0\phantom{-}\\
    \phantom{-}0\phantom{-} & \phantom{-}0\phantom{-} &\phantom{-}1\phantom{-}\\
             \end{pmatrix}\;,\quad
    \vec{r}=\mathbf{P}_z^{-1}(\xi)\vec{r}'\nonumber
 \end{equation}


%% file: var7-gamma3-english-step.tex
%

\section{Specific features of the time step divisible by\\ the solar day}
\label{sec:step}

\begin{figure}[htb]
        \centering
        \includegraphics[scale=1]{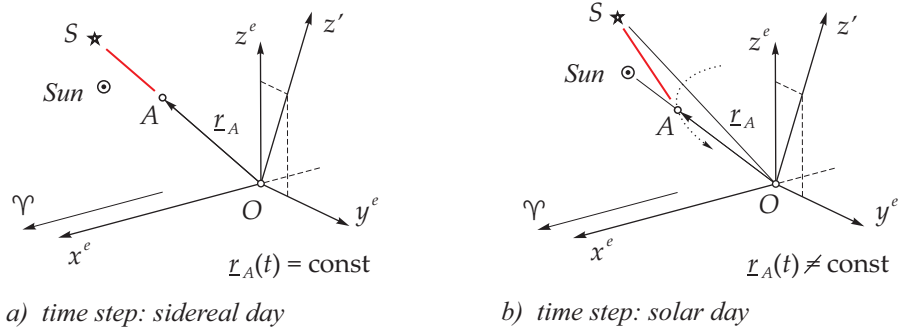}
        \caption{Dependence of character of the point $\mathbf{A}$ motion on the choice of the time series step ((a) or (b)). Axis $\mathbf{O}z'$ is the Earth's self-rotation axis. In the Earth's rotation about the Sun, axis $\mathbf{O}x^e$ always remains parallel to~\Aries}
        \label{fig:var7-gamma3-english-A-days}
     \end{figure}

What is very important in forming the procedure of measuring angles between the plumb line (local normal) and direction towards the star (fixed reference point) is to properly choose the time step. If the measurement series is formed with the interval divisible by the solar day, this means that the Earth (Observer $\mathbf{A}$) self-rotation about the axis with respect to stars is supplemented by rotation with angular velocity~$\omega_\text{year}$ about the same axis in the same direction. This is just the extra rotation of the Observer that must be taken into account in estimating the external perturbation periods. If the time series step is assumed to be equal to the sidereal day, the Observer will remain motionless with respect to stars. The ray drawn from point~$\mathbf{A}$ towards star~$\mathbf{S}$ will move only in the plane-parallel manner but will not rotate. If, however, the measurement time interval is equal to the solar day or multiple of it, the Observer will rotate with respect to stars with period $T_\text{year}$ and, hence, ray $\mathbf{AS}$ will rotate clockwise about point~$\mathbf{A}$. This is clearly shown in Fig.~\ref{fig:var7-gamma3-english-A-days}.

%% file: var7-gamma3-english-Luna.tex
%

\section{Moon's orbit}
\label{sec:Luna}

\begin{figure}[htb]
        \centering
        \includegraphics[scale=1]{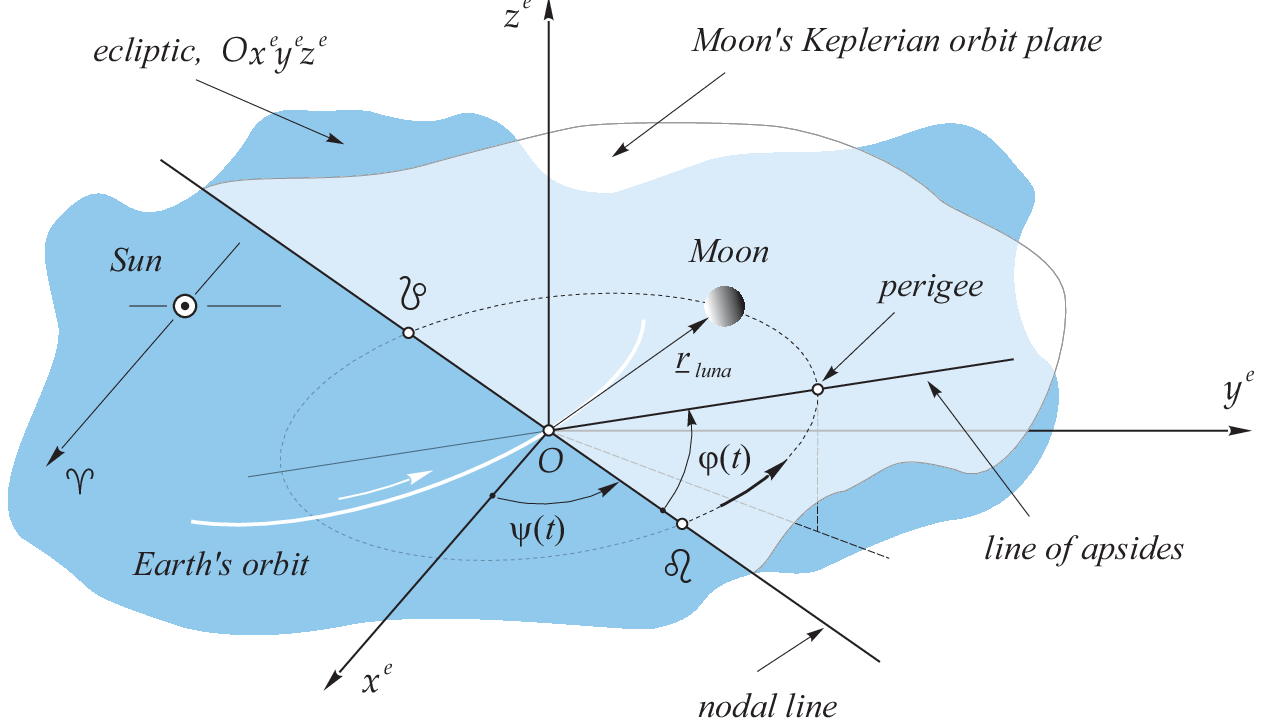}
        \caption{The Moon's orbit components in the $\mathbf{O}x^ey^ez^e$ coordinate system}
        \label{fig:var7-gamma3-english-OXeYeZe-Lune}
     \end{figure}

The Moon's orbit is a complex open spatial curve. The Moon's motion is considered with respect to fixed point $\mathbf{O}$ coinciding with the Earth's center of gravity (Fig.~\ref{fig:var7-gamma3-english-OXeYeZe-Lune}). The Moon's position in the $\mathbf{O}x^ey^ez^e$ coordinate system is specified by a combination of six cyclically unstable components of the orbit~\cite{1972:book:Kulikov}.
    \begin{equation}\label{eq:var7:lune6}
        \vec{r}_{luna}(t)=\vec{r}_{luna}\bigl(\: i(t),\psi(t),\varphi(t),e(t),a(t),t_\ast(t)\: \bigr)\;,
    \end{equation}
where $i(t)$ is the orbit inclination angle defined via the angle of intersection with ecliptic plane $\mathbf{O}x^ey^e$ and Moon's Keplerian trajectory plane; $\psi(t)$ is the longitude of the nodal
line (plane intersection line); in this case, the angle is measured from axis $\mathbf{O}x^e$ that is parallel to~\Aries\ at any time moment of the Earth's (point~$\mathbf{O}$) motion about the Sun; $\varphi(t)$ is the angle between the nodal line  and line of apsides; the Moon's trajectory ellipticity is defined by eccentricity~$e(t)$ and major semi-axis~$a(t)$; $t_\ast(t)$ is the time moment when the Moon passes the perigee.

For instance, relations for $\psi(t)$ and $\varphi(t)$ taken from~\cite{1972:book:Kulikov} for the 1900 epoch look as follows:
    \begin{eqnarray}\label{eq:var7:series-psi-phi}
       \begin{array}{rll}
        \psi(t)&=&\si{259\degree10\arcminute59\arcsecond\!{.}77} -
                     \si{1934\degree08\arcminute31\arcsecond\!{.}23}\cdot \tau +
                     \si{07\arcsecond\!{.}48}\cdot \tau^2 +
                     \si{0\arcsecond\!{.}0080}\cdot \tau^3\:,\\
        \varphi(t)&=&\si{\:\;75\degree\:\;8\arcminute46\arcsecond\!{.}61} +
                       \si{6003\degree10\arcminute33\arcsecond\!{.}75}\cdot \tau -
                       \si{44\arcsecond\!{.}65}\cdot \tau^2 -
                       \si{0\arcsecond\!{.}0530}\cdot \tau^3\:,\\
        &&\hfill\phantom{\int\limits^{.}}\tau(t)=(2415020 - t) / 36525\;,
        \end{array}
    \end{eqnarray}
where $\tau(t)$ is the time expressed in Julian centuries as a function of the current Julian
date~$t$. Time derivatives of those functions give us the nodal line and line of apsides rotation
periods:
    \begin{equation}\label{eq:var7:dpsidphi}
        T_{\psi}=\frac{2\pi}{\nicefrac{\dif{\psi}}{\dif{t}}}\approx-18.6\text{~years}\;\;,\qquad
        T_{\varphi}=\frac{2\pi}{\nicefrac{\dif{\varphi}}{\dif{t}}}\approx 6\text{~years}
    \end{equation}

Thus, according to the angular velocity summation rule, the perigee involved in those two rotations
about point~$\mathbf{O}$ in the $\mathbf{O}x^ey^ez^e$ coordinate system will have the following period:
    \begin{equation}\label{eq:var7:Tperigeum}
        T_\text{perigee}=\frac{T_{\psi}\cdot T_{\varphi}} {T_{\psi}+T_{\varphi}} \approx 8.85\text{~years.}
    \end{equation}

\paragraph{Moon's perigee mass.}

\begin{figure}[htb]
        \centering\includegraphics[scale=1]{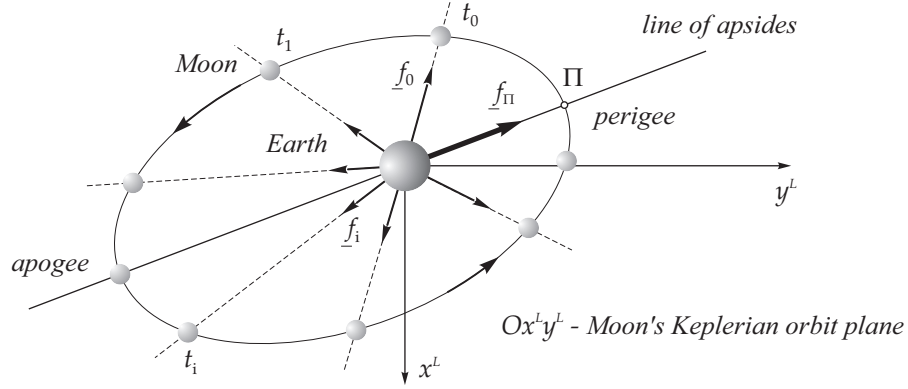}
        \caption{The resulting Moon's gravitational effect on the Earth during its revolution about
                            the Earth along the Keplerian orbit}
        \label{fig:var7-gamma3-english-Lune-28force-N}
     \end{figure}

Let us clarify how the Moon's perigee motion affects the direction and value of gravitational acceleration at point~$\mathbf{A}$ on the Earth's surface. Replace the Moon's gravitational effect on the Earth with the equivalent gravitational effect of a certain body located in the Moon's perigee. Let us derive this body mass from the Moon's gravitational effect on the motionless Earth (point~$\mathbf{O}$) during one cycle $T_{luna}\!\approx\!28\text{~days}$. Due to the axial symmetry and non-zero eccentricity, the resulting gravitational-type force will be directed towards the perigee (Fig.~\ref{fig:var7-gamma3-english-Lune-28force-N}).

Resulting force $\vec{f}_\Pi$ is an integral Moon's gravitational effect on the Earth during
the time interval equal to the period of the Moon's rotation about the~Earth:
    \begin{equation}\label{eq:var7:Fp}
        f_\Pi=|\vec{f}_\Pi|=\frac{1}{2\pi}\G M_{terra}^\ast M_{luna}
        \int\limits_0^{2\pi}\frac{\cos\alpha}{r(\alpha)^2}\;\dif{\alpha}\;,\quad
        M_{terra}^\ast=M_{terra}+M_{luna}\;,
    \end{equation}
where
    \begin{equation}\label{eq:var7:Fp-a}
        r(\alpha)=\frac{p}{1+e\cos\alpha}\;,\quad
        p=a(1-e^2)\;,\quad
        \alpha=\frac{2\pi}{\:T_{luna}\:}\:t\;.
    \end{equation}
Here $r(\alpha)$ is the Moon's focal radius as a function of angle~$\alpha$ measured counterclockwise
from the direction towards the perigee; $p$~is the focal parameter; $e$~is the eccentricity; $a$~is the major semi-axis. Integrating~\eqref{eq:var7:Fp},~obtain:
    \begin{equation}\label{eq:var7:Fp-final}
        f_\Pi=\G M_{terra}^\ast M_{luna}\frac{e}{p^2}\;.
    \end{equation}

Now we know the force $f_\Pi$ modulus and can derive an expression for a certain mass that can produce the necessary gravitational-type force effect. Below we refer to this mass as the Moon's~\emph{perigee~mass}. The law of the two-body gravitational interaction gives us the desired mass formula:
    \begin{equation}\label{eq:var7:mluna}
        m_\Pi= \frac{f_\Pi}{\G M_{terra}^\ast}\Pi_{luna}^2\;, \quad
        \Pi_{luna}=\frac{p}{1+e}\;,
    \end{equation}
where $\Pi_{luna}$ is the distance between the ellipse focus (point $\mathbf{O}$) and the perigee. Substituting \eqref{eq:var7:Fp-final} into \eqref{eq:var7:mluna}, obtain the \emph{perigee mass} formula:
    \begin{equation}\label{eq:var7:mlune-final}
        m_\Pi(e)= M_{lune} \frac{e}{(1+e)^2}\;,\;\;\;\quad
        m_\Pi(0) = 0\;,\quad
        m_\Pi(e_{luna}) \approx \frac{1}{20} M_{luna}\;.
    \end{equation}

Thus, we have defined the Moon's perigee mass as a function of eccentricity of the Moon's Keplerian orbit. As it was noted earlier, variations in the Moon's orbit components are of the cyclic character; therefore, in the first approximation we assume eccentricity~$e(t)$ to be a harmonic function whose period is equal to the time of the perigee revolution about the Earth's center~\cite{1972:book:Kulikov}:
\begin{equation}\label{eq:var7:e}
        e(t)=\overline{e}+\frac{1}{2}\bigl(e_{max}-e_{min}\bigr)
        \sin\left(\frac{2\pi}{T_\text{perigee}}\:t\right)\;,\quad
        \overline{e}=const\;.
    \end{equation}

Introduction of fictitious perigee mass $m_\Pi(e)$ allowed us to exclude the "fast" component of the Moon's motion and to consider only the Moon's perigee~$\boldsymbol{\Pi}$ motion. The perigee mass position in the fixed coordinate system $\mathbf{O}x^ey^ez^e$ is defined by vector:
\begin{equation}\label{eq:var7:luna5}
        \vec{r}_\Pi(t)=\overbrace{\Pi_{luna}\bigl(e,a\bigr)}^{a(1-e)}\cdot
            \mathbf{P}_z\bigl(\psi(t)\bigr)\!\cdot\!
            \mathbf{P}_x\bigl(i(t)\bigr)\!\cdot\!
            \mathbf{P}_z\bigl(\varphi(t)\bigr)\!\cdot\! \vec{e}_1
    \end{equation}


%% file: var7-gamma3-english-plumb-line.tex
%

\section{Plumb line and Moon's perigee}

\begin{figure}[htb]
        \centering
        \includegraphics[scale=1]{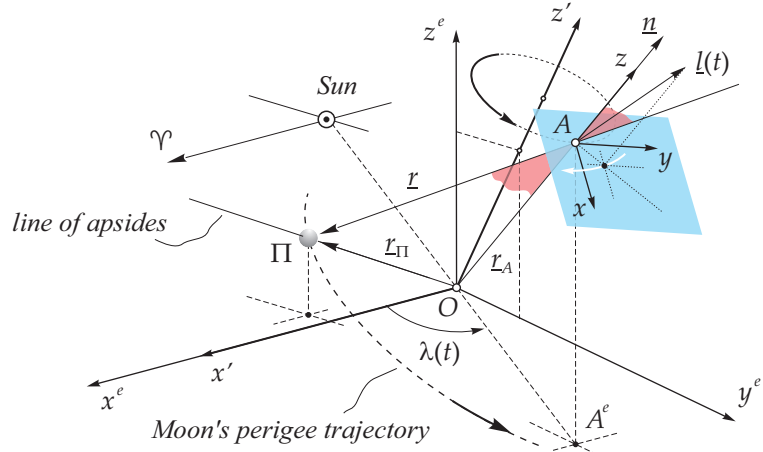}
        \caption{The Moon's perigee $\boldsymbol{\Pi}$ trajectory and Observer $\mathbf{A}$ coordinate system $\mathbf{A}xyz$ on the rotating Earth's surface in the fixed coordinate system~$\mathbf{O}x^ey^ez^e$}
        \label{fig:var7-gamma3-english-OXeYeZe-Axyz}
     \end{figure}

Let us reveal how angle $\gamma(t)$ depends on relative positions of the Moon's perigee $\boldsymbol{\Pi}$ and Observer $\mathbf{A}$ on the rotating Earth surface (Fig.~\ref{fig:var7-gamma3-english-OXeYeZe-Axyz}).

    \begin{figure}[htb]
        \centering
        \includegraphics[scale=1]{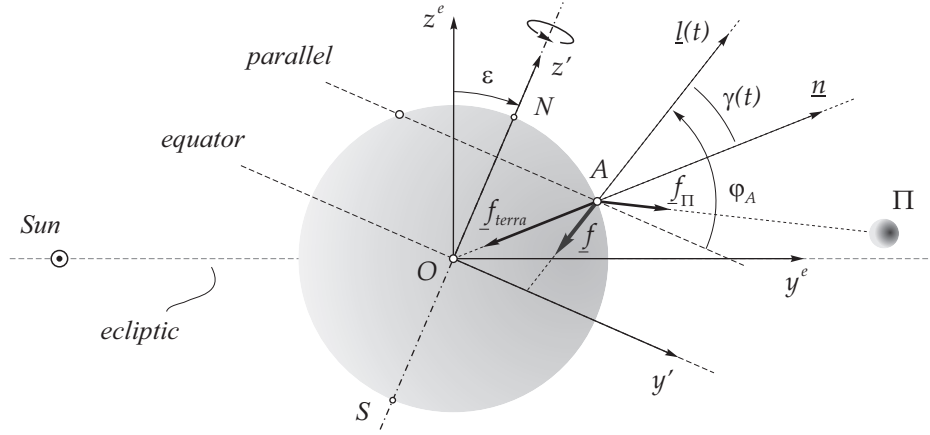}
        \caption{Gravitational forces acting upon the proofmass at point~$\mathbf{A}$ at the moment
                        when the Sun, Earth and Moon's perigee mass are in the~$\textbf{O}y^ez^e$}
        \label{fig:var7-gamma3-english-OXeYeZe-force}
     \end{figure}

In the coordinate system~$\mathbf{O}x^ey^ez^e$, point $\mathbf{A}$ (Observer) and point $\boldsymbol{\Pi}$ (Moon's perigee mass) rotate along their trajectories about point~$\mathbf{O}$ (the Earth's center of gravity) in the same direction (counterclockwise) with different angular velocities. Proofmass~$m_A$ at point~$\mathbf{A}$ (Fig.~\ref{fig:var7-gamma3-english-OXeYeZe-force}) undergoes two gravitational forces: from the Earth and from Moon's perigee~mass.
    \begin{equation}\label{eq:var7:F}
            \vec{f}_{terra}(t)=\G\;m_A M_{terra}\;
            \frac{\vec{r}_A}{|\vec{r}_A|^3}\;,\quad
            \vec{f}_\Pi(t)=\G\;m_A m_\Pi\;
            \frac{\vec{r}_\Pi-\vec{r}_A}{|\vec{r}_\Pi-\vec{r}_A|^3}\;,
    \end{equation}
where $\vec{r}_\Pi(t)$, $\vec{r}_A(t)$ are the vectors of the Moon's perigee \eqref{eq:var7:luna5} and Observer \eqref{eq:var7:rA}, respectively. Designate as $\vec{f}(t)$ the sum of forces acting at point~$\mathbf{A}$:
    \begin{equation}\label{eq:var7:f2}
            \vec{f}(t) = \vec{f}_{terra}(t) + \vec{f}_\Pi(t)\;,
    \end{equation}
In this case, force $\vec{f}(t)$ in the moving coordinate system $\mathbf{A}xyz$ (Fig.~\ref{fig:var7-gamma3-english-OXeYeZe-Axyz}) looks like:
    \begin{equation}\label{eq:var7:fAxyz}
        \vec{f}_A(t)=\mathbf{P}_A(t) \!\cdot\! \vec{f}(t)\;,\quad
        |\vec{f}_A(t)|=|\vec{f}(t)|\;,
    \end{equation}
where
    \begin{equation}\label{eq:var7:PA}
        \mathbf{P}_A(t)=
        \mathbf{P}_x\left(-\varepsilon\right) \!\cdot\!
        \mathbf{P}_z\left(\lambda(t)\right)  \!\cdot\!
        \mathbf{P}_y\left(\frac{\pi}{2}-\varphi_A\!\right)\;.
    \end{equation}

\begin{figure}[htb]
        \centering
        \includegraphics[scale=1]{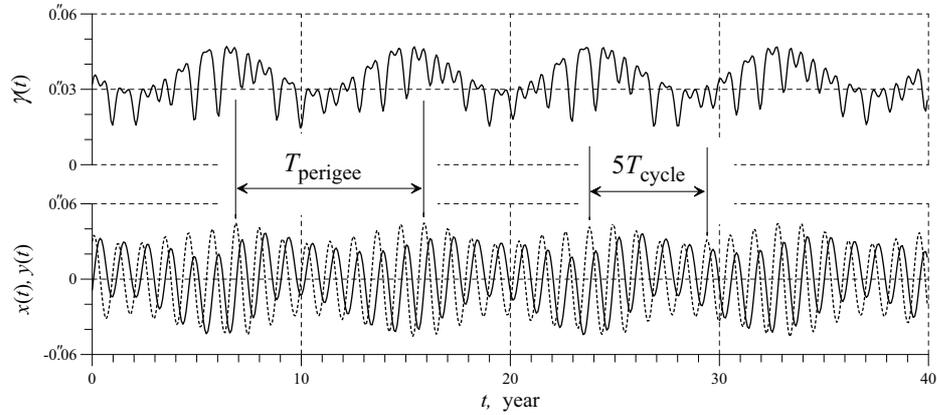}
        \caption{Time dependence of angle $\gamma(t)$ and its components $x(t)$, $y(t)$}
        \label{fig:var7-gamma3-english-graphics-Agamma}
     \end{figure}

\begin{figure}[htb]
        \centering
        \includegraphics[scale=1]{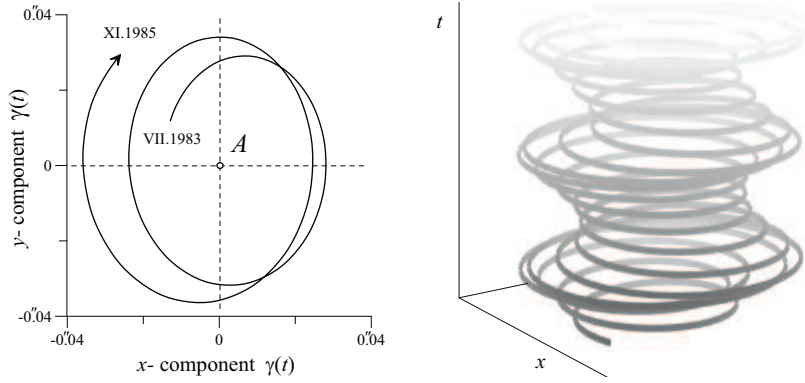}
        \caption{A fragment of the vector $\vec{l}(t)$ vertex trajectory projection on plane ~$\mathbf{A}xy$
            and its time scanning ($t$ is the third coordinate)}
        \label{fig:var7-gamma3-english-graphics-Apoloid-2D-3D}
     \end{figure}

The calculations showed that in this coordinate system vector $\vec{l}(t)$ vertex moves cyclically clockwise about normal~$\vec{n}$ with period $T_\text{cycle}\!\approx\!411.8$~days. This process is illustrated in Figs.~\ref{fig:var7-gamma3-english-graphics-Agamma} and \ref{fig:var7-gamma3-english-graphics-Apoloid-2D-3D}.

Period $T_\text{cycle}\!\approx\!411.8\text{~days}$ in the rotating coordinate system $\mathbf{A}xyz$ is a sum of two rotations: the Observer's (point $\mathbf{A}$) rotation about axis $\mathbf{O}z'$ with the period equal to that of the Earth's rotation about the Sun and the Moon's perigee rotation about the Earth's center of gravity. The long-period ($T_\text{perigee}\!\approx\!8.85$ years) component of the observed process (Fig.~\ref{fig:var7-gamma3-english-graphics-Agamma}) is determined only by cyclic variations in the Moon's elliptic trajectory eccentricity and major semi-axis.

Naturally, continuous and non-random variation in the gravitational situation at the observation point on the Earth's surface affects the spatial location of the plumb line or line normal to the artificial horizon~\cite{1988:PATENT:N1718632, 1988:PATENT:N1832962}.


%% file: var7-gamma3-english-comparison.tex
%

\section{Comparison of calculations and observations}
\label{sec:comparison}

Spatial location of the zero point (plumb line) of the astronomical instrument is determined by instantaneous gravitational field configuration and makes the instrument gravitationally independent. Therefore, high angle measurement accuracy adds to the instrument properties of high-sensitive gravimeter able to detect minor long-period gravity field perturbations at the observation point in measuring the angles.

Let us compare the plumb line (local normal) behavior under the influence of the Moon's perigee mass with the results of long-term observation of variations in the angle between the local normal and direction to the stars. These observations miss the nutation-precession perturbations of the Earth's rotation axis caused by the known external forces.

The above-considered natural effect of the Moon's perigee mass on the local normal indicates that the plumb line oscillations with the amplitude of $\pm0^{''}\!\!.06$ and period $T_\text{cycle}\!\approx\!411.8$~days are similar to the Chandler's wobble. This makes it possible to assume that the Chandler's wobble is of the gravitational nature and depends on the Moon's perigee mass.

\begin{figure}[htb]
        \centering
        \includegraphics[scale=1]{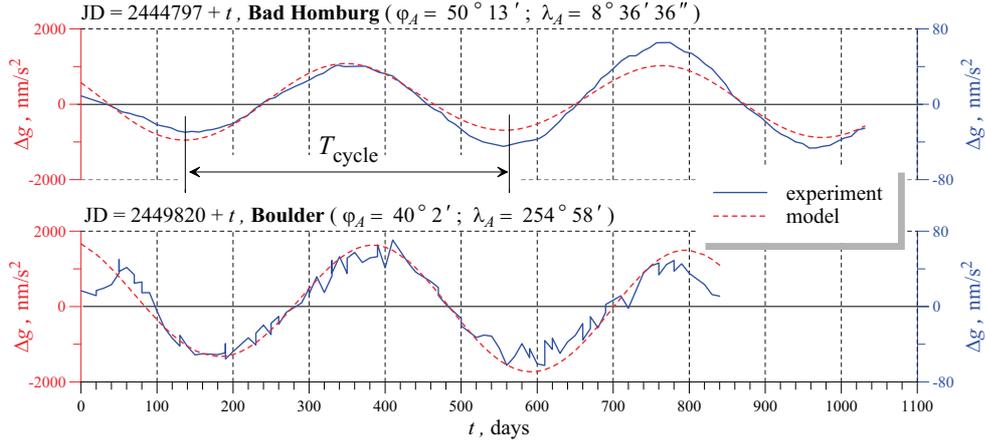}
        \caption{Comparison of calculations and observations of the gravitational acceleration $\Delta\g$ variations (observations were performed at the \emph{Bad Homburg} and \emph{Boulder} stations). The series are asynchronous, the origin of each series is given by the Julian date~JD}
        \label{fig:var7-gamma3-english-graphics-Adg-BadHomburg-Boulder}
     \end{figure}

Results of the process numerical simulation fits well the results of long-term observations
of the Earth's gravitational field normal component $\Delta\g$ performed at the
\emph{Bad~Homburg}~\cite{1998:FGS} and \emph{Boulder}~\cite{1998:GeoRL-v25-p393:Boulder} stations. The Fig.~\ref{fig:var7-gamma3-english-graphics-Adg-BadHomburg-Boulder} curves show that even when the station longitudes differ by about $246$ degrees, the gravity field variation period remains constant and equal to the Chandler's period ($T_\text{cycle}\!\approx\!411.8$~days).

\begin{figure}[htb]
        \centering\includegraphics[scale=1]{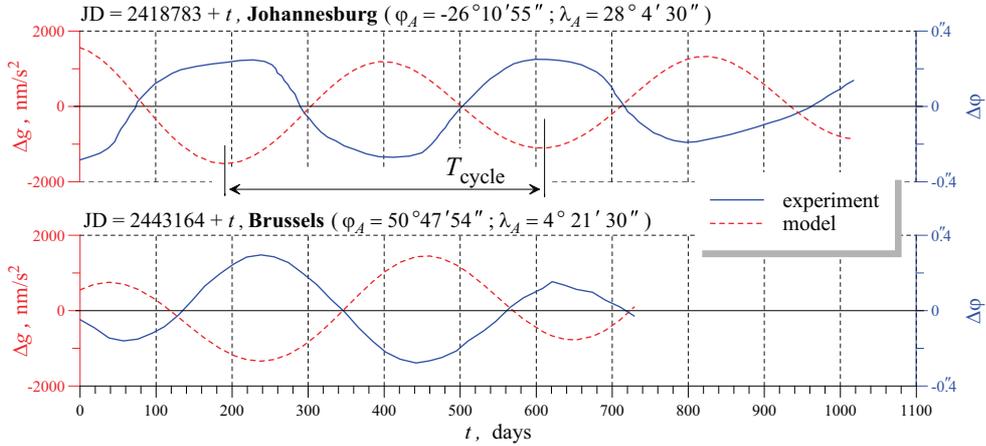}
        \caption{Comparison of calculated variations in gravitational acceleration $\Delta\g$ with observation data on "latitude variation" $\Delta\varphi$ obtained at stations \emph{Johannesburg} and \emph{Brussels}. The series are asynchronous, the origin of each series is given by the Julian date~JD}
        \label{fig:var7-gamma3-english-graphics-Adg-Johannesburg-Brussels147D}
     \end{figure}

Observations of "latitude variation" $\Delta\varphi$ and variation in gravity field normal component $\Delta\g$ at stations \emph{Johannesburg}~\cite{1956:book:Johannesburg} and \emph{Brussels}~\cite{1982:book:Podobed} are presented in Fig.~\ref{fig:var7-gamma3-english-graphics-Adg-Johannesburg-Brussels147D}. The curves behavior confirms our assertion that the latitude variation and oscillation of the metering instrument plumb line are of the same (gravitational) character.

It is important to note that the observed so-called "latitude variation" $\Delta\varphi$ (to say more exactly, variation in the field line (local normal) angle of deviation from the direction to the star) and variation in the gravitational acceleration $\Delta\g$ normal component have the same period and are strictly anti-phase; this is just that should take place in the Earth-Moon gravitational interaction.

Optical astrometry detects \emph{something} as the zenith distance variations with the Chandler's period and amplitude of $0^{''}\!\!.2$ to $0^{''}\!\!.3$ angular seconds. The metering instrument local normal (plumb line) varies with the same period and phase but with the amplitude of $\pm0^{''}\!\!.06$. This allows us to confidently assume that the detected \emph{something} referred to as the Chandler's wobble is of the gravitational character.

In recent decades, zenith distances began to be measured by radio-astronomical methods and tools instead of optical instruments; this transition did not induce uncertainty in the results of the detected process analysis and did not distort its amplitude and frequency characteristics. The results of observations analysis are almost identical. This confirms that in both cases the same physical phenomenon is detected.

Identity of observations obtained by different methods and fundamentally different instruments shows that a phenomenon of one and the same character is observed and detected, and this character is gravitational;
just this changes the spatial location of the Earth's rotation axis and the Earth itself. This occurs due gravitational forces of the Moon's perigee mass acting on the Earth's equatorial mass excess. In other words, in both cases instruments detect the effect of nutation momentum produced by the Moon's perigee mass upon the Earth, which is not taken into account in the existing precession-nutation theory. Therefore, the Chandler's wobble consists in variation in the Earth's rotation axes motion jointly with the Earth with respect to stars rather than in the rotation axis motion within the Earth.

Chandler's wobbles $\pm0^{''}\!\!.2$ in amplitude were detected in experiments with a laser gyroscope~\cite{Schreiber2011, Schreiber2013}; this confirms our conclusion about the forced character of the Earth's rotation axis motion jointly with the Earth caused by the gravitational momentum induced
by the Moon's perigee~mass.


%% file: var7-gamma3-english-IERS.tex
%

\section{IERS time series}
\label{sec:IERS}

\begin{figure}[htb]
        \centering\includegraphics[scale=1.2]{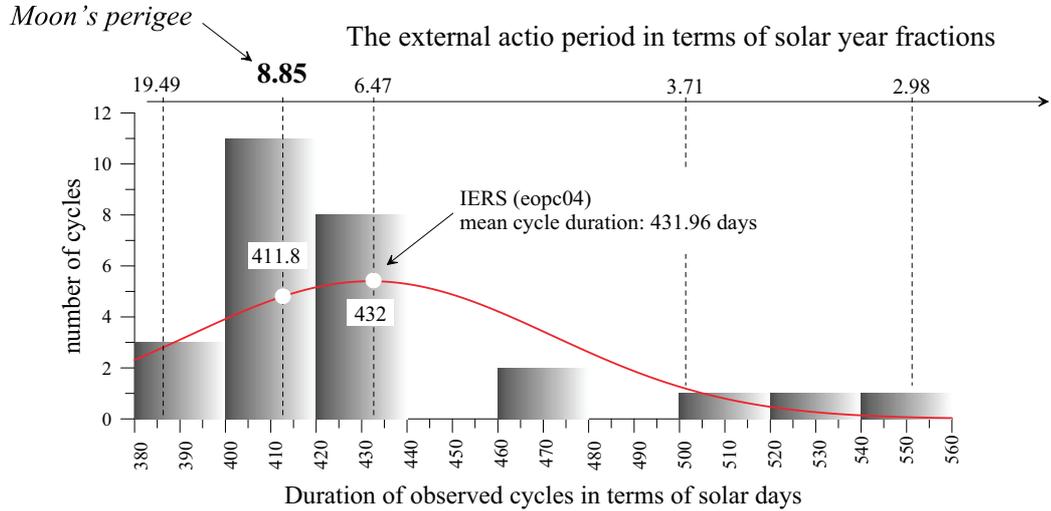}
        \caption{Distribution of ($2\pi$) cycles over duration in the "residual motion" of the
                            Earth's rotation axis within the Earth}
        \label{fig:var7-gamma3-english-graphics-histogram}
     \end{figure}

The standard approach to processing observations on variations in celestial body zenith distances is that the maximum attention is given to statistical methods, the physical meaning being almost fully ignored. Analysis of observations only by statistical methods, ignoring the detected phenomenon physics, often pushes the investigation into the metaphysics territory. The histogram given in Fig.~\ref{fig:var7-gamma3-english-graphics-histogram} shows the results of our
analysis of~IERS\footnote{International Earth Rotation and Reference Systems Service, \url{http://www.iers.org}\\ \hspace*{7.1mm}Earth Orientation Center, \url{http://hpiers.obspm.fr/eop-pc}}. series. The histogram shows that the period of $\approx\!411.8$~days occurs most frequently.

Now, under the assumption that the process with this period is a result of external impact, it is necessary to take into account the Earth's extra annual rotation involved in the solar day definition~\eqref{eq:var7:OmegaTerra}. The Observer's self-rotation with the~$T_\text{year}$ period changes the periods of external perturbing factors detected in observations. For instance, the Moon's perigee mass rotates counterclockwise with period $T_\text{perigee}\!=\!8.85$~years ($3232.46$~days) about the Earth's center of gravity and perturbs the Earth's gravity field. In the scope of classical mechanics, the period of variation in the plumb line spatial location caused by this perturbation, which is detected by the Earth's Observer~$\mathbf{A}$, may be defined as follows:
    \begin{equation}\label{eq:var7:TA1}
        T_\text{cycle}=\frac{T_\text{perigee} \cdot  T_\text{year}}{T_\text{perigee} - T_\text{year}}=
            \frac{8.85 \cdot 1}{8.85 - 1}\approx 1.13\,\text{year}\;\;\;\text{или}\;\;
            \approx 411.8\,\text{days}\;.
    \end{equation}

If the time scale step is not multiple of the solar day, the same period of the
plumb line gravitational perturbation will be different:
    \begin{equation}\label{eq:var7:TA1}
        T_\text{cycle}=\frac{T_\text{perigee} \cdot T_\text{days}}{T_\text{perigee} - T_\text{days}}=
            \frac{3232.46 \cdot 1}{3232.46 - 1}\approx 1.00031\,\text{days}\;.
    \end{equation}

A somewhat similar period is considered in the paper of A.S.\,Vasilyev's~\cite{1952:article:Vasiliev} who studied variations in the star zenith distances. The distinctive feature of his work is that he formed time series for the analysis with the step not multiple of solar day. In our opinion, he in fact observed the same process as S.\,Chandler.

Consider again the Fig.~\ref{fig:var7-gamma3-english-graphics-histogram} histogram.
The lower (linear) scale presents the cycle durations ignoring the Observer’s self-rotation with
period $T_\text{year}$. The top (non-linear) scale is adequate to the lower scale of cycle durations but is corrected for the Observer's self-rotation with the~$T_\text{year}$ period. For the period of about $411.8$~days observed on the Earth (we call it secondary), the external perturbing process period (primary period) ranges from $8$ to $11$~years. The only possible source of this perturbation is the perigee Moon~\eqref{eq:var7:mlune-final} whose period of rotation about the Earth is $T_\text{perigee}$. Therefore, we can affirm that the period of about $411.8$~days is of the natural origin but is secondary~\emph{per~se}.


%% file: var7-gamma3-english-ABCD.tex
%

\section{Example of plumb line gravitational dependence}
\label{sec:ABCD}

\begin{figure}[htb]
        \centering
        \includegraphics[scale=1]{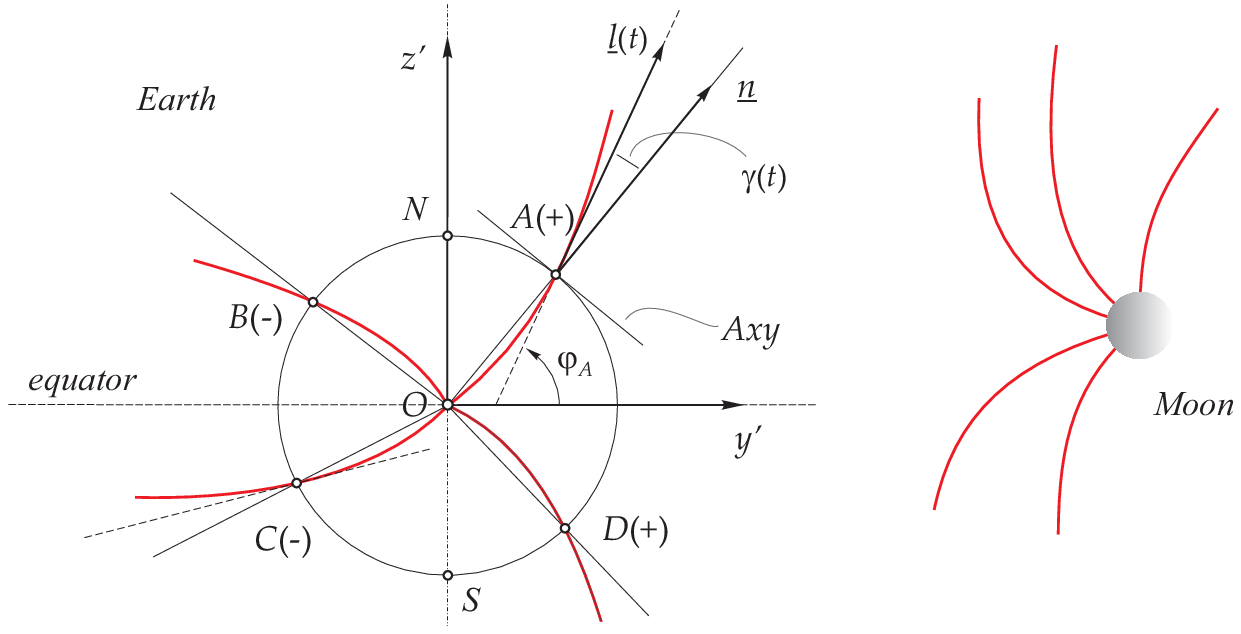}
        \caption{The Moon's effect on plumb line $\vec{l}(t)$ deviation from normal $\vec{n}$ to 
        plane $\mathbf{A}xy$ tangential to the Earth's surface at point~$\mathbf{A}$}
        \label{fig:var7-gamma3-english-ABCD-latitudes}
     \end{figure}

Consider the Moon's role in variation in spatial location of the gravity field line tangential (plumb line) at points $\mathbf{A}$, $\mathbf{B}$, $\mathbf{C}$, $\mathbf{D}$ on the rigid Earth's surface. Let the Earth-Moon distance be fixed at time moments when point~$\mathbf{A}$ and Moon are in the plane $\mathbf{O}y'z'$, point~$\mathbf{A}$ being on the side facing the Moon (Fig.~\ref{fig:var7-gamma3-english-ABCD-latitudes}). At the same time moments, let us fix angle~$\gamma(t)$ between plumb line~$\vec{l}(t)$ and normal~$\vec{n}$ to tangential $\mathbf{A}xy$ at point~$\mathbf{A}$.

The presence of the Moon's gravitational mass induces apparent latitude increase at points $\mathbf{A}$ and $\mathbf{D}$ and apparent latitude decrease at points $\mathbf{B}$ and $\mathbf{C}$.

As the Moon approaches the Earth, plumb line deviation $\od{\gamma}{t}>0$ induces the effect of the geographical (astronomical) latitude\footnote{The astronomical (or geographical) local latitude is the angle between the equator plane and plumb line. In this case, the plumb line is defined only by the Earth's gravity field.} increase at points $\mathbf{A}$ and $\mathbf{D}$ and decrease at points $\mathbf{C}$ and $\mathbf{B}$ where~$\od{\gamma}{t}<0$. Thus, when star zenith distances are observed simultaneously at the same meridian but on different sides of the equator, plumb line~$\vec{l}(t)$ deviations are in-phase.


%% file: var7-gamma3-english-conclusion.tex
%

\section{Summary}
\label{sec:conclusion}

It happened so that results of observations of local latitude variations performed in 19th century were understandably perceived as a consequence of the Earth's rotation axis motion within the Earth. In this case, the fact that the so-called "observed latitude behavior" may result from minor long-period gravitational action of the Moon on the Earth was almost fully ignored, as well as the fact that interpretation of the observations analysis results depends on the choice of the observation series time step. This moment may be regarded as a bifurcation point in forming a paradigm that has determined the character of further investigation of the Chandler's wobble physical nature. In summary, let us formulate the main result:

\begin{quote}
    \emph{We have established the existence of external gravitational momentum acting upon the Earth from the Moon’s perigee mass, which has not been involved in the existing theory of the Earth’s precession-nutation.}
\end{quote}

Consideration of the Moon's perigee mass in the precession-nutation theory will eliminate such \emph{entity} as the Chandler's motion of poles. This, in its turn, will cause revision of the Earth's model and rotation theory, modification of corrections to the Universal Time System, satellite navigation systems (GPS, GLONASS), geodesic measurements and physical experiments. Only gravitational dependence of the local normal spatial position and possible illusions of the local latitude variation will remain (Fig.~\ref{fig:var7-gamma3-english-ABCD-latitudes}).
